\documentclass[preprint2,flushrt]{aastex}
\usepackage{natbib}
\usepackage{amssymb,amsmath}

\shorttitle{Gamma-ray flare from the Crab}
\shortauthors{Buehler et al.}
\begin{document}

\title{Gamma-ray Activity in the Crab Nebula: \\The Exceptional Flare of April 2011}

\author{
R.~Buehler\altaffilmark{1,2},
J.~D.~Scargle\altaffilmark{3,4}, 
R.~D.~Blandford\altaffilmark{1,5},
L.~Baldini\altaffilmark{6}, 
M.~G.~Baring\altaffilmark{7}, 
A.~Belfiore\altaffilmark{8,9,10}, 
E.~Charles\altaffilmark{1}, 
J.~Chiang\altaffilmark{1},  
F.~D'Ammando\altaffilmark{11,12}, 
C.~D.~Dermer\altaffilmark{13}, 
S.~Funk\altaffilmark{1}, 
J.~E.~Grove\altaffilmark{13}, 
A.~K.~Harding\altaffilmark{14}, 
E.~Hays\altaffilmark{14}, 
M.~Kerr\altaffilmark{1}, 
F.~Massaro\altaffilmark{1}, 
M.~N.~Mazziotta\altaffilmark{15}, 
R.~W.~Romani\altaffilmark{1}, 
P.~M.~Saz~Parkinson\altaffilmark{8}, 
A.~F.~Tennant\altaffilmark{16},
M.~C.~Weisskopf\altaffilmark{16}
}
\altaffiltext{1}{W. W. Hansen Experimental Physics Laboratory, Kavli Institute for Particle Astrophysics and Cosmology, Department of Physics and SLAC National Accelerator Laboratory, Stanford University, Stanford, CA 94305, USA}
\altaffiltext{2}{email: buehler@stanford.edu}
\altaffiltext{3}{Space Sciences Division, NASA Ames Research Center, Moffett Field, CA 94035-1000, USA}
\altaffiltext{4}{email: Jeffrey.D.Scargle@nasa.gov}
\altaffiltext{5}{email: rdb3@stanford.edu}
\altaffiltext{6}{Istituto Nazionale di Fisica Nucleare, Sezione di Pisa, I-56127 Pisa, Italy}
\altaffiltext{7}{Rice University, Department of Physics and Astronomy, MS-108, P. O. Box 1892, Houston, TX 77251, USA}
\altaffiltext{8}{Santa Cruz Institute for Particle Physics, Department of Physics and Department of Astronomy and Astrophysics, University of California at Santa Cruz, Santa Cruz, CA 95064, USA}
\altaffiltext{9}{Universit\`a degli Studi di Pavia, 27100 Pavia, Italy}
\altaffiltext{10}{INAF-Istituto di Astrofisica Spaziale e Fisica Cosmica, I-20133 Milano, Italy}
\altaffiltext{11}{IASF Palermo, 90146 Palermo, Italy}
\altaffiltext{12}{INAF-IRA Bologna, I-40129 Bologna, Italy}
\altaffiltext{13}{Space Science Division, Naval Research Laboratory, Washington, DC 20375-5352, USA}
\altaffiltext{14}{NASA Goddard Space Flight Center, Greenbelt, MD 20771, USA}
\altaffiltext{15}{Istituto Nazionale di Fisica Nucleare, Sezione di Bari, 70126 Bari, Italy}
\altaffiltext{16}{NASA, Marshall Space Flight Center, Huntsville, AL 35812}

%\tableofcontents

\begin{abstract}
The Large Area Telescope on board the \textit{\textit{Fermi}} satellite 
observed a gamma-ray flare in the Crab nebula 
lasting for approximately nine days in April of 2011. The source, which at optical wavelengths has a size of $\approx$11 ly  across,
doubled its gamma-ray flux within eight hours.  The peak photon flux 
was $(186 \pm 6) \times 10^{-7}$ cm$^{-2}$ s$^{-1}$ 
above 100 MeV, which corresponds to a 30-fold increase compared to the average value.
During the flare, a new component emerged in the spectral
energy distribution, which peaked at an energy of (375 $\pm$ 26) MeV
at flare maximum. The observations imply that the emission  region was
likely relativistically beamed toward us and that variations in its motion are
responsible for the observed spectral variability.

\end{abstract}

\section{Introduction}
The Crab nebula is the remnant of a 
supernova observed in 1054 AD. 
The explosion left behind a rotating neutron star 
emitting electromagnetic radiation 
pulsed at the rotation period,
that powers a wind of relativistic particles.
These particles interact with the remnant gas and magnetic field, 
causing the nebula to glow brightly at all
wavelengths, predominantly by 
synchrotron radiation. 
The spectral energy distribution (SED) of the nebula is, 
accordingly, 
dominated by a synchrotron component 
extending from radio wavelengths into the gamma-ray band \citep{hester_crab_2008}. Above 450 MeV a second component emerges, attributed to inverse-Compton  
scattering
by the same relativistic particles \citep{gould_1965,dejager_1992,atoyan_fluxes_1996}.
The angular size of the Crab nebula is $\approx$0.1$^\circ$
in the optical and smaller at higher energies. This corresponds to 3.5 pc, or 11 ly, at its estimated distance of 2 kpc \citep{trimble_distance_1973}.

Today, the pulsar and nebula 
(henceforth referred to together as the Crab) 
are considered prime examples of non-thermal sources in the Universe and serve 
as a laboratory for relativistic plasma physics. 
New puzzles for our understanding of the Crab have been posed by the detection of three bright gamma-ray flares by the AGILE satellite and the Large Area Telescope (LAT) on board the \textit{Fermi} satellite between 2007 and 2010 \citep{abdo_crab_2011,tavani_crab_2011}. 
During these flares the unpulsed component of the gamma-ray flux increased by a factor of {$\approx$10} on time scales as short as 12 hours \citep{balbo_crab_2011}, 
while the period and flux of the pulsed component remained stable. 

More recently, 
in April of 2011 the LAT detected a fourth flare,
three times brighter than any of the previous ones \citep{buehler_crab_2011,hays_crab_2011}. 
The flare was swiftly confirmed by the AGILE satellite \citep{striani_crab_2011}. Observations at lower energies, in particular by the Chandra X-ray observatory, have not yet revealed any variability
correlated with the gamma-ray flares \citep{tennant_2011}. However, analysis of these observations 
is ongoing and will be discussed elsewhere. 
Here we present the LAT gamma-ray results obtained during the flare, together with a broader analysis of 
the first 35 months of Crab observations by \textit{Fermi}.

\section{The Large Area Telescope and data analysis}
The LAT is a pair-conversion telescope, sensitive to 
gamma rays with energies greater than 20 MeV. 
It has a large field of view {($\approx$2.4 sr)} and images the full sky 
every three hours.
The angular resolution of the LAT 
varies with photon energy. The 68\% containment radius ranges  
from approximately 6$^{\circ}$ at 70 MeV to 0.2$^{\circ}$ above 10 GeV \citep{atwood_2009}. 
LAT scientific observations began in August 2008. 
We analyzed data taken within 20$^{\circ}$ of the Crab in the first 35 months of observations (MJD 54683--55728). 
The average spectral properties of the Crab were derived from the first 33 months of observations (MJD 54683--55664), excluding the April 2011 flare. 

Fluxes and spectra were obtained 
by maximizing the likelihood 
of source models using unbinned {\tt gtlike} from the 
\textit{Fermi} Science Tools 9-23-01.
The models included  all sources in the second LAT source catalog within 20$^{\circ}$ of the Crab position \citep{latcatalog}
plus models for the Galactic and isotropic diffuse emission ({\tt gal\_2yearp7v6\_v0, iso\_p7v6source}). The parameters left free to vary in the likelihood fit 
for 33-month average spectra were the spectral parameters of the Crab, the normalization of the diffuse components and a power-law spectral index for scaling the Galactic diffuse emission model. For analysis on shorter time scales only the isotropic diffuse normalization was varied along with the Crab spectral properties. All other parameters were fixed to the 33-month average maximum likelihood values. The Sun was included in the source model when it was within 20$^{\circ}$ of the Crab. The solar spectra during the two passages in front of the Crab were taken from \citet{abdo_fermisun_2011}. The Moon was not included in the source model as its gamma-ray contamination was found to be negligible for observations presented here.

We used the P7{\_}V6{\_}SOURCE instrument response functions without in-flight PSF corrections, selecting photon events
between 70 MeV and 300 GeV. 
Compared to the more typical 100 MeV threshold 
this choice leads to additional systematic errors 
due to increased dispersion in the photon energy reconstruction. 
The overall systematic flux error is energy dependent: 
it amounts to 30\% at 70 MeV and decreases to 10\% above 10 GeV \citep{rando_2011}. 
The dominant part of this systematic error is related to the overall flux normalization. It is caused by uncertainties in the effective area determination and of the overall normalization of energy scale.

The stability of the LAT instrument over time was tested for all time scales addressed in this publication using the Vela and Geminga pulsars, which are found to be stable in flux \citep{abdo_fermivela_2010,abdo_fermigeminga_2010}. Their flux variations were $<$10\%, yielding an upper limit on the variations of the systematic errors with time. A more detailed study of the systematic uncertainties is currently being prepared within the LAT collaboration for publication.

The pulsar phase was assigned to the detected gamma rays based on a high-time-resolution pulsar ephemeris. To obtain the latter
we extracted 400 pulsar times-of-arrival (TOAs) from LAT photons collected from MJD
54684-55668 with a typical uncertainty of $\sim$100$\mu$s \citep{ray_2011}.  Using
{\tt tempo2} \citep{hobbs_2006}, we fit a timing solution to these TOAs with a typical residual
of 108$\mu$s, or about $3\times10^{-3}$ of the pulsar period.  To
obtain these white residuals, we modeled the pulsar timing noise
using the method of \citet{hobbs_2004}, with 20 harmonically related sinusoidal terms. The ephemeris parameter file, and light curves and spectra shown in this publication are publicly available online\footnote{{http://www-glast.stanford.edu/pub\_data/691/}}.

\begin{figure}[t!]
\centering
\noindent\includegraphics[width=0.46\textwidth]{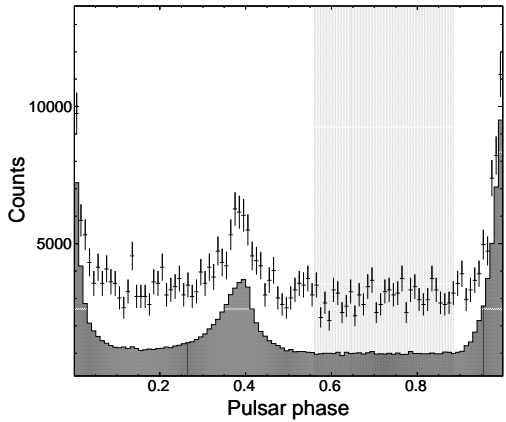}
\caption{Phase profile of gamma rays above 70 MeV within 3$^{\circ}$ of the Crab Pulsar for the first 33 months of \textit{Fermi} observations (black histogram) 
and during the April 2011 gamma-ray flare between MJD 55663.70--55671.02 (black markers
with error bars). The gray region indicates the adopted off-pulse 
interval where the emission is dominated by the nebula.
The flare phase profile has been multiplied by a factor 59, such that the excess above the off-pulse counts is the same as for the 33 months of observations. This demonstrates that the flare is a phase-independent flux increase.}
\label{phase}
\end{figure}

\begin{figure}[t!]
\centering
\noindent\includegraphics[width=0.46\textwidth]{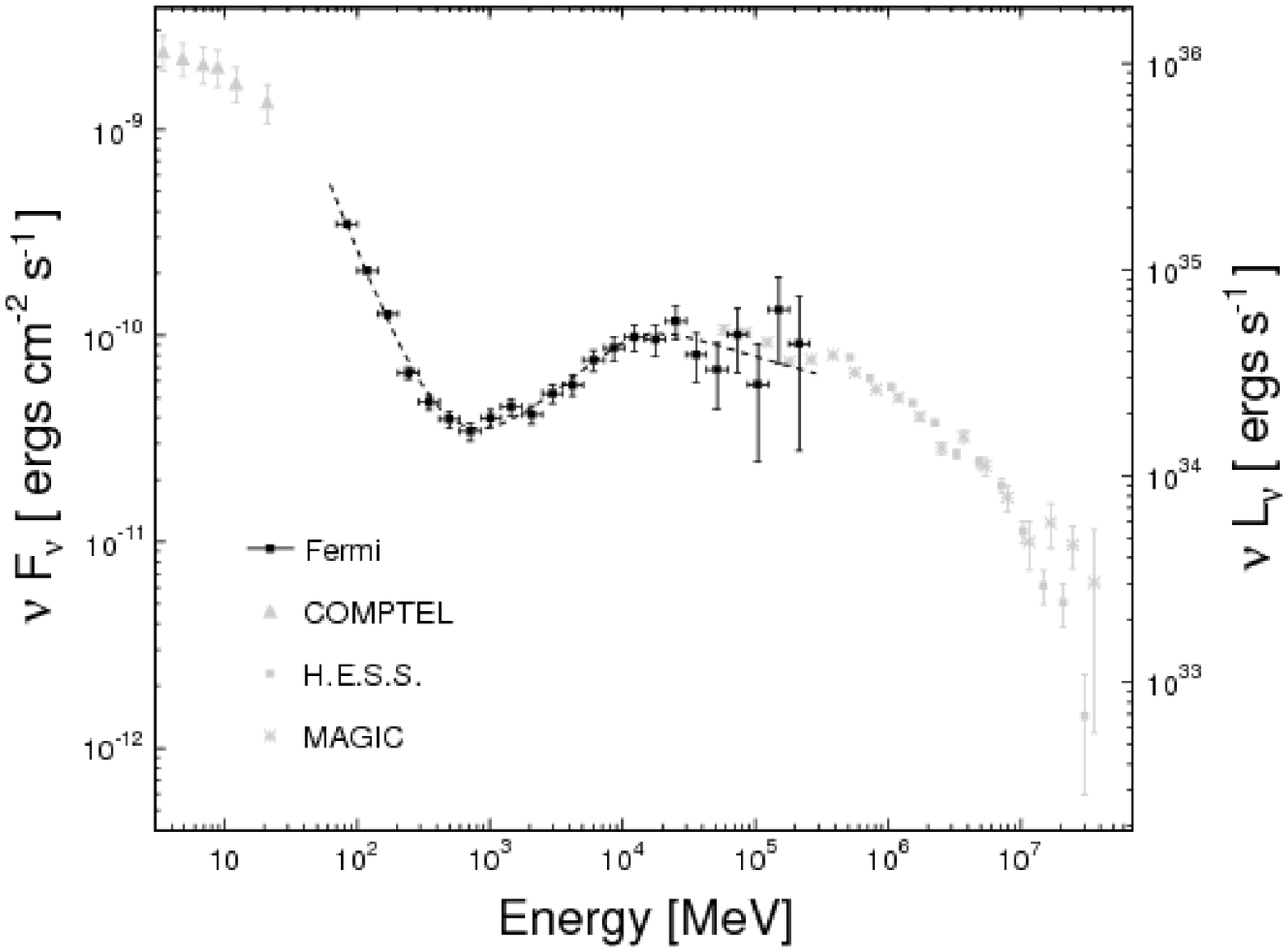}
\caption{Spectral energy distribution for 
the Crab nebula averaged over the first 33 months of \textit{Fermi} observations. The axis on the right side indicates the isotropic luminosity. Also shown are data from COMPTEL in the soft gamma-ray band \citep{kuiper_2001} and very high energy gamma-ray measurements from Cherenkov telescopes  \citep{aharonian_hess_2006,albert_magicnebula_2011}. The dashed line shows the maximum likelihood model in the parametrization described in the text.}
\label{avspec}
\end{figure}

\section{Time-Average Energy Spectra}

The Crab appears to the LAT as a point source, even at the highest photon energies.  
To separate nebular gamma-ray emission from that of the pulsar we apply
a cut on the pulsar phase. 
The phased count rate of the pulsar 
with its double-peaked structure is shown in Figure \ref{phase}. 
The pulsar dominates the phase-averaged gamma-ray flux, 
but its flux in the off-pulse interval from 0.56 to 0.88
 is negligible \citep{abdo_fermi_2010}. 
It is in this interval that we measure the properties of the nebula.

The LAT detects the nebula in the energy range between the
the high-energy end of the synchrotron 
and the low-energy end of the inverse-Compton 
components of the SED. The average nebular spectrum measured during the first 33 months
of \textit{Fermi} observations is shown in Figure \ref{avspec}. We fitted it as the sum of synchrotron and inverse-Compton components.
 The differential photon spectrum, $\Phi(E)$, of the synchrotron component was parametrized with a power law 
 \begin{equation} 
 \Phi_{S}(E) =  \frac{F_{S}~(\gamma_{S}-1)}{(100~\mathrm{MeV})^{1-\gamma_{S}}}~E^{-\gamma_{S}} ,
 \end{equation}
 with an integral photon flux above 100 MeV of {$F_{S} = (6.1 \pm 0.2) \times 10^{-7}$ cm$^{-2}$ s$^{-1}$} and a soft photon index of {$\gamma_{S} = 3.59  \pm 0.07$}. 
  
 The spectrum of the inverse-Compton component 
 softens significantly with respect to a power-law at higher energies, as expected from measurements at very high energy (VHE) gamma rays \citep{aharonian_hess_2006,albert_magicnebula_2011,abdo_fermi_2010}. The spectrum was parametrized by a smoothly broken power law with a curvature index of {$\beta$ = 0.2}:
 \begin{equation}
 \begin{split}
 \Phi_{I}(E) = \Phi_{I,0} &\left(\frac{E}{100~\mathrm{MeV}}\right)^{-\gamma_{I,1}} \times \\
 &\left(1 + \left(\frac{E}{E_b}\right)^{\frac{\gamma_{I,2} - \gamma_{I,1}}{\beta}} \right)^{-\beta}. 
 \end{split}
 \end{equation}
 \noindent 
 The spectrum has a photon index of {$\gamma_{I,1} = $1.48 $\pm$ 0.07} at low energies and softens to {$\gamma_{I,2} =$ 2.19 $\pm$ 0.17} above a break energy of {$E_b = $ (13.9 $\pm$ 5.8) GeV}. 
 The flux normalization is {$\Phi_{I,0} = (5.6 \pm 1.4) \times 10^{-10}$ cm$^{-2}$ s$^{-1}$} MeV$^{-1}$ and the integral flux above 100 MeV is $F_{I} = (1.1 \pm 0.1) \times 10^{-7}$ cm$^{-2}$ s$^{-1}$. 

The averaged pulsar spectrum in the first 33 months of observations was measured in the on-pulse after accounting for the nebula emission. We parametrized the pulsar spectrum with a power-law function with a super-exponential cutoff
\begin{equation}
\Phi_{P}(E) = \Phi_{P,0} \left(\frac{E}{1~\mathrm{GeV}}\right)^{-\gamma_{P}} e^{-\left(\frac{E}{E_{P,c}}\right)^\kappa} .
\label{superexp}
\end{equation}
\noindent 
The best-fit value for the spectral index is  $\gamma_{P}=1.59 \pm 0.01$, for the break energy $E_{P,c}= (504 \pm 63)$ MeV and for the curvature index $\kappa=0.43 \pm 0.01$. The normalization is given by {$\Phi_{P,0} = (8.1 \pm 0.5) \times 10^{-10}$ cm$^{-2}$ s$^{-1}$} MeV$^{-1}$. The integral flux above 100 MeV is $F_{P} = (20.4 \pm 0.1) \times 10^{-7}$ cm$^{-2}$ s$^{-1}$, in agreement with the value measured after eight months of observations \citep{abdo_fermi_2010}. 

The pulsed emission from the Crab has recently been detected between $\approx$25-400 GeV. This emission is constrained to pulses which are narrower by a factor of approximately two compared to the LAT energy range. Pulsed emission at these energies challenges current pulsar models \citep{aliu_magicpulsar_2008,aliu_2011,aleksic_2011}. A more detailed spectral analysis of the Crab pulsar in the \textit{Fermi}-LAT energy range and its connection to the pulsed VHE emission is currently being performed and will be published elsewhere.

\section{Temporal flux variations}

\begin{figure*}[]
\centering
\noindent\includegraphics[width=0.96\textwidth]{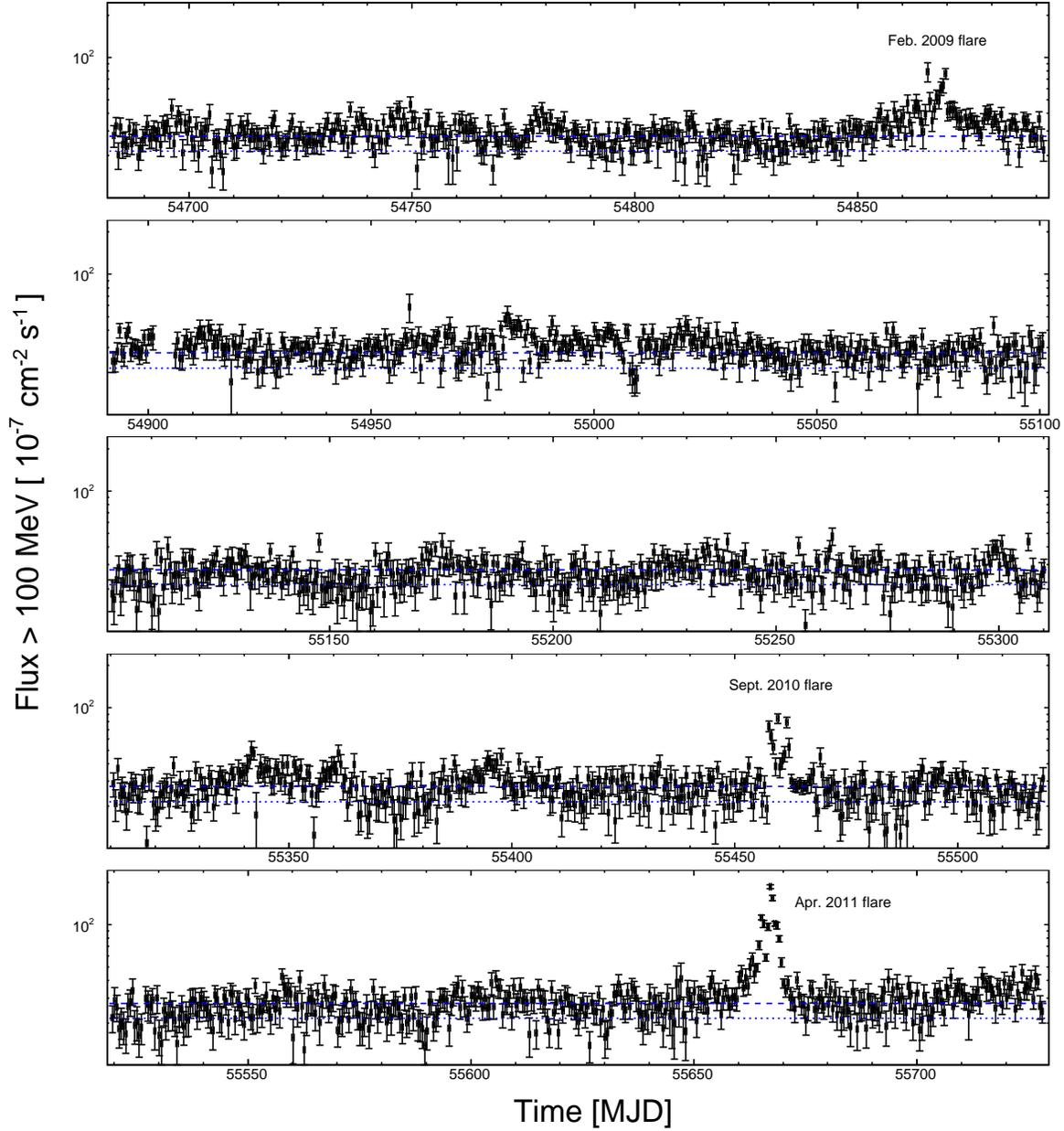}
\caption{Integral flux above 100 MeV from the Crab as a function of time for the first 35 months of \textit{Fermi} observations. The time binning is 12 hours. The dotted blue line indicates the sum of the 33-month average fluxes from the inverse-Compton nebula and the pulsar. The dashed blue line shows the average flux of the synchrotron nebula summed to the latter. Fluxes are shown on a log scale. The three flares detected by the \emph{Fermi} LAT are indicated. A fourth flare was detected in October 2007 (MJD 54380) by AGILE \citep{tavani_crab_2011}.}
\label{lcall}
\end{figure*}

In order to probe the flux variation over time, the flux 
from the Crab was evaluated in 12-hour time intervals. 
The combined pulsar and nebular spectrum was modeled as a power-law in energy, 
as further spectral features are not resolvable on these 
short time scales for typical observed fluxes. 

\begin{figure}[t!]
\centering
\noindent\includegraphics[width=0.46\textwidth]{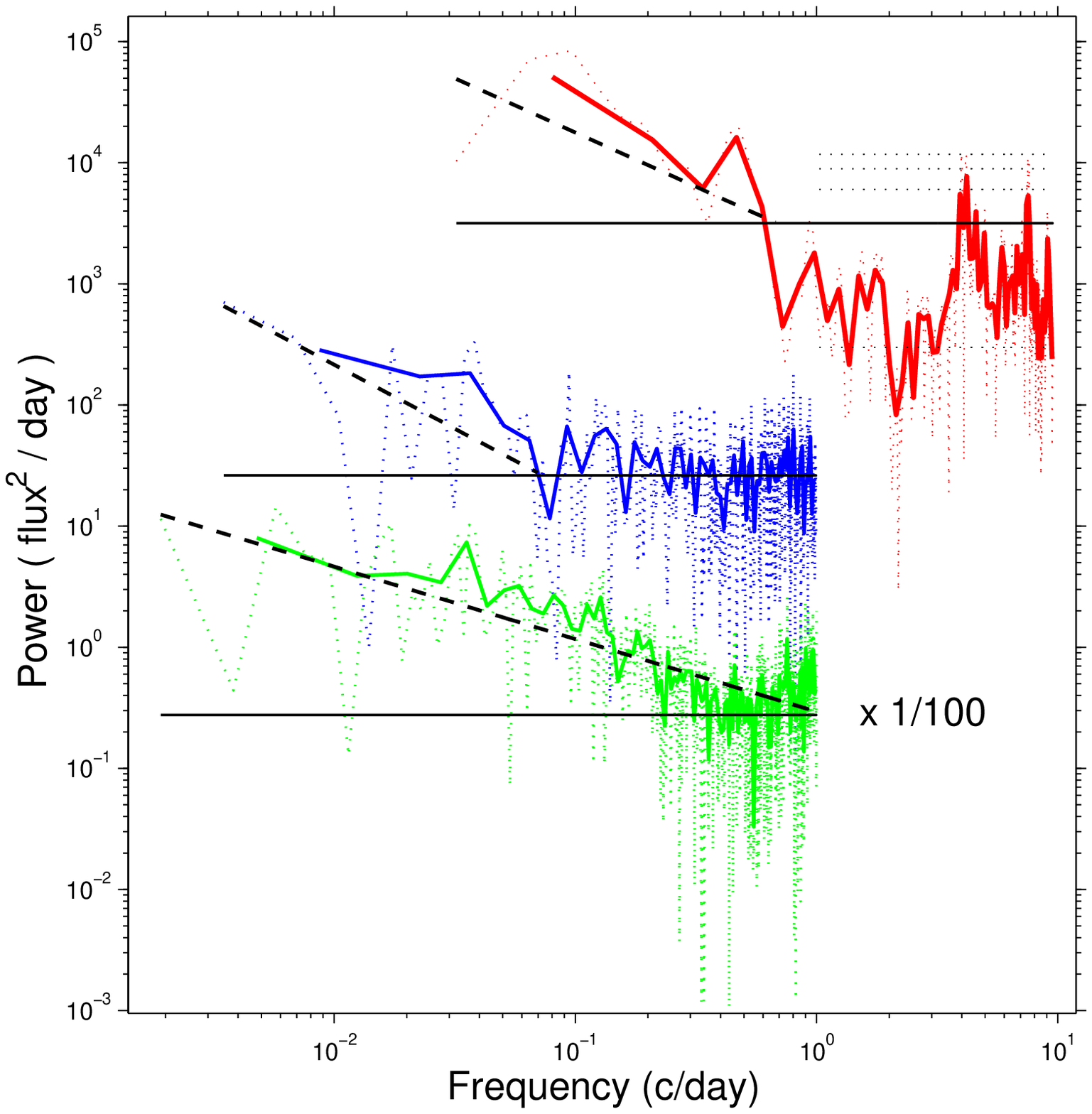}
\caption{Crab nebula Fourier power density spectrum (PDS), calculated from the light curve of the first 35 months of \textit{Fermi} observations shown in Figure \ref{lcall}. The PDS of the full time interval is shown by the solid green line (scaled down by 1/100 for better visibility). The PDS of the low activity period between MJD 54884--55457 is shown by the solid blue line. The PDS of the April 2011 flare is indicated by the solid red line and was calculated from the light curve shown in Figure \ref{lcflare}. A smoothing with a running average of four bins was applied to all spectra. The PDSs obtained before smoothing are shown in colored dotted lines. Black lines show the best fit function of a power-law function (dashed) plus a constant white noise component (solid) for the unsmoothed spectra. The best fit spectral indices are given in the text. Dotted black lines indicate the $\pm 1 \sigma$, the $+2 \sigma$ and $+3 \sigma$ confidence intervals derived from white noise simulations for the April 2011 flare PDS.} 
\label{spectrum}
\end{figure}

The resulting light curve is shown in Figure \ref{lcall}, where the three flares previously reported by the \textit{Fermi-LAT} collaboration are indicated. Variability is seen over the whole 35-month period. This is also apparent in the Fourier power density spectrum (PDS) shown in Figure \ref{spectrum}, which is approximately described by a power law with a spectral index of {$\approx$0.9}. The PDS was derived using the fast Fourier transform for evenly spaced data, interpolating over the short time interval MJD 54901--54906, during which the Crab was not observed by \textit{Fermi}. We verified that results are 
essentially indistinguishable from power spectra computed using the techniques for unevenly sampled time series described later in this section. The PDS shows significant power above the noise level on time scales from years to weeks. Variations on these time scales are also present outside of the flaring periods. This can be seen in the PDS of the time interval MJD 54884--55457 also shown in Figure \ref{spectrum}, where the nebula did not show large variations in flux. The spectral index of the PDS during this time is approximately 1.0.

There are compelling reasons to believe that the phase-averaged Crab pulsar and the inverse-Compton component flux are constant in the LAT energy range \citep{dejager_1992,atoyan_fluxes_1996}, or at least that they vary only very slowly as the pulsar spins down, and that any rapid variability is attributable to the synchrotron component of the nebula. 
This was confirmed observationally on monthly time scales 
for the first 24 months of \textit{Fermi} observations \citep{abdo_crab_2011}. 
To test this assumption 
on the 12-hour time scales studied here, 
we measured the flux in the on-pulse interval, 
subtracting the nebula flux  measured in the off-pulse interval
for each time bin. 
The pulsar flux was found to be stable within 20{\%}. No significant variations were found.
The inverse-Compton component of the nebula is too faint to be consistently 
detected in 12-hour time windows. Therefore the variability of this
component cannot be quantified from our analysis on these time scales.
However, significant flux variations of the inverse-Compton component 
on short time scales are strongly disfavored 
theoretically.
The observed absence of variations on monthly time scales
is consistent with constancy on the shorter scales
of interest here.

On 2011 April 9 the flux of the Crab increased dramatically. 
Once the source reached flux levels comparable to the 2009 and 2010 flares the \textit{Fermi} satellite was commanded to switch to 
a pointed observation targeting the Crab (MJD 55663.70--55671.02). 
In this mode the exposure toward the source increased by a factor of about four compared to the standard all-sky monitoring. 
During these observations the Crab erupted to a peak flux 
of $(186 \pm 6) \times 10^{-7}$ cm$^{-2}$ s$^{-1}$ above 
100 MeV during the 12-hour period centered at MJD 55667.14, as shown Figure \ref{lcall}. 
This corresponds to a flux increase of a factor 7 compared to the average total flux from the Crab
and a factor of 30 compared to the flux from the nebula. We verified that the flare is indeed positionally coincident with the 
Crab nebula, with a best-fit localization of {R.A. = 83.65$^\circ$} {Dec. = 21.98$^\circ$} (J2000),
and an error radius of 0.04$^\circ$ at 95\% confidence.

\begin{figure*}[t!]
\centering
\noindent\includegraphics[width=0.96\textwidth]{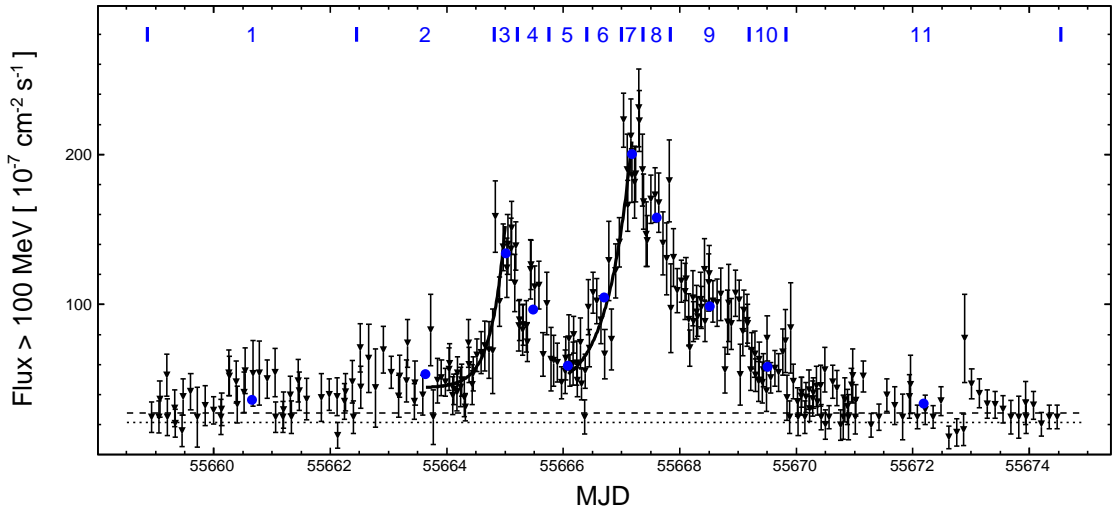}
\caption{Integral flux above 100 MeV as a function of time during the 2011 April Crab flare. The light curve is binned into equal exposure bins during times with no Earth occultation, with a mean bin duration of nine minutes. The dotted line indicates the sum of the 33-month average fluxes from the inverse-Compton nebula and the pulsar. The dashed line shows the flux of the average synchrotron nebula summed to the latter. The solid black lines show the best fit of a model consisting of a constant plus an exponential function at the rise of both sub-flares (see text). The blue vertical lines indicate the intervals of each Bayesian Block during which the flux remains constant within statistical uncertainties. The time windows are enumerated at the top of the panel. The corresponding flux is shown by the blue marker below each number. The SED for each of the time windows is shown in Figure \ref{specevol}. }
\label{lcflare}
\end{figure*}

The high gamma-ray flux during the flare, combined with the increased exposure of the pointed observations allowed us to study the flux evolution down to time scales below the $\approx$1.5 hour orbital period of \textit{Fermi}. On such time scales occultation of the source by the Earth needs to be taken into account. The time binning was therefore adjusted such that only time periods during which the Crab is visible to the LAT are used. These visibility windows were further split into bins of equal exposure, yielding a mean bin  duration of nine minutes. The evolution of the flux during the flare in this binning is shown in Figure \ref{lcflare}. The flare lasted for approximately 9 days and is composed of two sub-flares, peaking around MJD 55665 and MJD 55667. During both, the flux increased rapidly, reaching its maximum value within approximately one day. A second rise or ``shoulder'' is observed during the decaying phase of both sub-flares. Whether this is coincidental cannot be assessed on the  basis of these two events alone. If these shoulders are not interpreted as additional flares of lower amplitude, the decay time is approximately 1.5 and 3 days for the first and second sub-flare, respectively. 

To investigate the significance of peaks on smaller time scales in the flare light curve, we decomposed it into periods compatible with a constant flux. The time windows are referred to as ``Bayesian Blocks'' (BBs). They were determined finding the partition which maximizes the sum of the cost function assigned to each BB \citep{scargle_bayesianb_1998}. The cost function was set to the logarithm of the maximum likelihood for the constant flux hypothesis. The algorithm to determine the optimal partition is described by \citet{jackson_2005}. The BB-binned light curve is shown in Figure \ref{lcflare}. It is statistically compatible with the original light curve ({$\chi^2_r/ndf$ = 257/232}). This implies that flux variations within each BB cannot be distinguished with confidence from a locally constant flux. The shortest BBs are detected at the maximum of both sub-flares and have  durations of $\approx$9 hours. 

In order to measure the rate of flux increase at the rising edges of the sub-flares we parametrized them with an exponential function plus a constant background. The best-fit functions are shown in Figure \ref{lcflare}. The time ranges over which the fits were performed were defined by the centers of the BBs before and at the maximum of each sub-flare. The resulting doubling time is {4.0 $\pm$ 1.0} hours and {7.0 $\pm$ 1.6} hours for the first and second sub-flare, respectively. As these values depend on the somewhat arbitrarily chosen parametrization and fit ranges, we conservatively estimate that the doubling time scale in both sub-flares is t$_{d} \lesssim 8$ hours.

\begin{figure*}[t]
\centering
\noindent\includegraphics[width=0.96\textwidth]{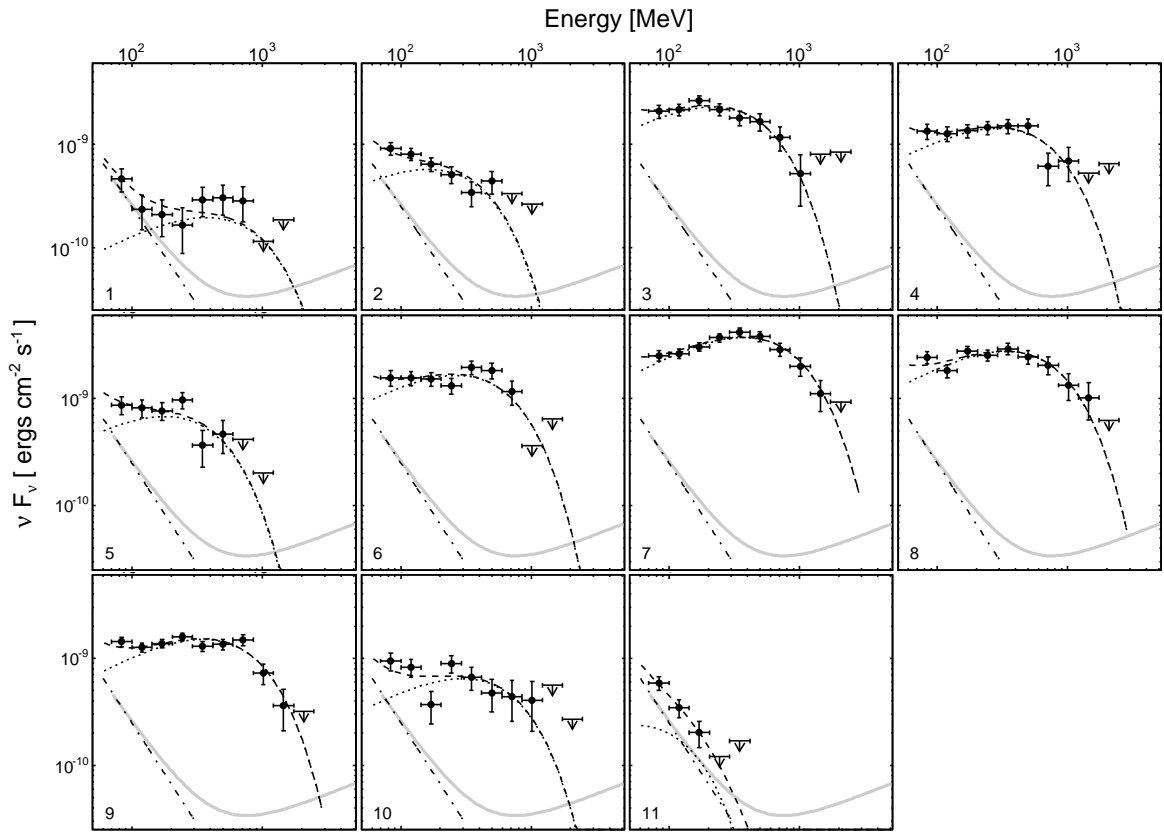}
\caption{Spectral Energy Distribution evolution during the April 2011 Crab flare. Arrows indicate 95\% confidence flux upper limits. The time windows are indicated in the bottom left corner of each panel and correspond to the ones indicated in Figure \ref{lcflare}. The dotted line shows the SED of the flaring component, the dot-dashed line the constant background from the synchrotron nebula, and the dashed line is the sum of both components (see text). The average Crab nebular spectrum in the first 33 months of \textit{Fermi} observations is also shown in gray for comparison. }
\label{specevol}
\end{figure*}

The PDS of the April 2011 flare is shown in Figure \ref{spectrum}. It was obtained by computing the Fourier transform of the autocorrelation function using an algorithm for unevenly sampled data \citep{edelson_1988}. The PDS can be described by a power law of index {$\approx$1.1} and reaches the noise floor at a frequency of $\approx$0.6 cycles per day. The doubling time of the corresponding sinusoidal component is $\approx$10 hours, in agreement with the expectation from the measured doubling times of the flares.

The pulsar flux remained unchanged during the flare, with an average flux above 100 MeV of $F_{P} = (21.7 \pm 1.1) \times 10^{-7}$ cm$^{-2}$ s$^{-1}$ during the main part of the flare (MJD 55663.70--55671.02). The flux increase is phase-independent. This is illustrated in Figure \ref{phase}, where the phasogram during the main flare period is shown. The peaks in the on-pulse interval remain at the same position. We also searched for periodicities other than the Crab pulsar with the time-differencing technique \citep{atwood_2006}, applying the event-weighting technique described in \citet{bickel_2008}. We scanned the frequency range 0.1--256 Hz, allowing for a possible spindown up to twice the value of the Crab pulsar. No significant signal was found besides the pulsar, which was detected with a significance $> 5.5 \sigma$. Finally, we searched for photon clumping on time scales shorter than the $\approx$10 min time binning by applying a Bayesian Block analysis on the single photon arrival times, with no significant detection.

\subsection{Spectral evolution during the flare}

In order to measure the energy spectrum during the flare, and its evolution with time, the data must be averaged in time intervals long enough to ensure adequate photon statistics, but short enough to provide adequate temporal resolution. The 11 bins of approximately constant flux, derived from the BB analysis, provide a reasonable compromise between these two constraints. 

The SEDs for each of the time bins are shown in Figure \ref{specevol}, after subtracting the steady emission from the pulsar and the inverse-Compton component of the nebula. It can be clearly seen that a new spectral component emerges from the synchrotron nebula during the flare, moving into the \textit{Fermi} energy range as the flare evolves. Its flux reaches a maximum between MJD 55666.997--55667.366 (frame 7); during this period the peak in the SED is clearly detected at {$E_{peak}$ = (375 $\pm$ 26) MeV}. 

It is difficult to parametrize the spectral shape of the flaring component due to the likely contamination by background flux from the synchrotron nebula not related to the flare. The determination of the latter is degenerate with the measurement of the flare component, as only the summed flux is measured. To break this degeneracy we proceeded under the following assumptions: \begin{enumerate}
\item The spectrum of the synchrotron nebula during the flare can be described by a power-law function and does not vary in time.
\item The spectrum of the flaring component can be described by a power-law with an exponential cutoff (equation \ref{superexp} with $\kappa=1$). While the cutoff energy and normalization of the spectrum vary, the spectral index remains constant during the flare.
\end{enumerate}
We derived the spectral index of the flaring component and the spectrum of the background synchrotron component in a composite likelihood fit to all the time windows displayed in Figure \ref{lcflare}, simultaneously measuring the energy cutoff and flux normalization evolution of the flaring component in each of the time windows. For this we used the {\tt composite likelihood 2} part of the \textit{Fermi} Science Tools.  

\begin{figure}[t]
\centering
\noindent\includegraphics[width=0.46\textwidth]{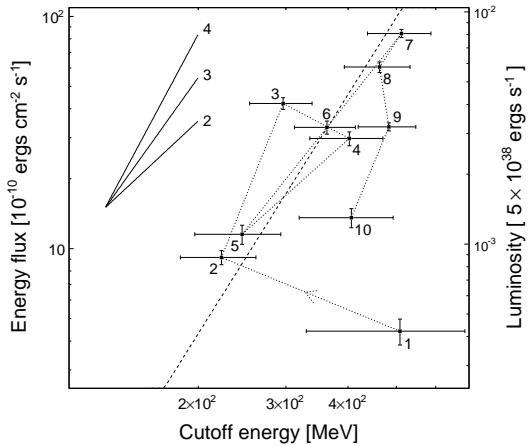}
\caption{Total energy flux above 100 MeV, $f_{F}$, as a function of cutoff energy E$_c$ for the flaring component of the April 2011 flare. The values correspond to the parametrizations shown in frames 1--10 of Figure \ref{specevol} (the values obtained from frame Nr 11 are not included, as no significant spectral curvature was detected in his time interval, allowing no robust determination of E$_c$). The number next to each marker denotes the corresponding frame.  The axis on the right hand side indicates the apparent luminosity in units of the pulsar spin-down power of $5 \times 10^{38}$ ergs s$^{-1}$ \citep{hester_crab_2008}. The numbered solid lines indicated the slope of the corresponding power-law dependency $f_{F} \sim {E_{F,c}}^\alpha$. The dotted line marks the best fit function in this parametrization with {$\alpha$ = 3.42 $\pm$ 0.86}. }
\label{corr}
\end{figure}

The best-fit values for the background synchrotron nebula during the flare period are {$F_{S}$ = (5.4 $\pm$ 5.2) 10$^{-7}$ cm$^{-2}$ s$^{-1}$} and {$\gamma_{S}$ = 3.9 $\pm$ 1.3}, consistent with the average value measured during the first 33 months of observation. The spectral index of the flaring component is measured to be {$\gamma_{F}$ = 1.27 $\pm$ 0.12}. The best-fit values for $E_{F,c}$ and the energy flux above 100 MeV, $f_{F}$, are shown in Figure \ref{corr}. 

This simple parametrization yields a good description of the flare evolution, as can be seen in Figure \ref{specevol}. The data are therefore compatible with the interpretation that a new spectral component of a fixed spectral shape evolves during the flare. As shown in Figure \ref{corr} the cutoff energy of the flaring component varies significantly, having a statistical probability of being constant of only 0.04\%. However, while $f_{F}$ varies by more than an order of magnitude, the cutoff energy varies only by approximately a factor of two. The relationship between both quantities can be described approximately by a power-law function $f_{F} \sim {E_{F,c}}^\alpha$, with best-fit index of {$\alpha$ = 3.42 $\pm$ 0.86}. 

\section{Discussion}

A year after first being reported, the gamma-ray flares from the Crab remain enigmatic. Where within the nebula does the emission come from? What produces the flux variations? How were the emitting particles accelerated? How are the flares related to the variability observed on yearly and monthly time scales? Although several ideas have been proposed, no certain answers can be given today. The observations presented here give us the most precise look into the flare phenomenon to date, during the brightest outburst detected so far. We will proceed to discuss some of the implications and challenges posed by these observations.

One striking property of the Crab nebula flares is their rapid flux variations, doubling within {t$_{d} < 8$ hours} at the rise of the 2011 April flare. Causality arguments imply that the emission region is compact, with a length {$L < c t_{d} \approx 2.8 \times 10^{-4}$ pc}. The emitted isotropic power at the peak of the flare of $\approx4 \times 10^{36}$ erg s$^{-1}$ corresponds to $\approx$1\% of the total spin-down power of the pulsar, the ultimate energy source of the nebula.  It is difficult to explain how this energy is focused into such a small emission volume. The focusing is generally easier to explain when the emission site is closer to the pulsar. The absence of pulsation in the flare signal implies that the emission region is at least located outside the light cylinder of the pulsar. Another possibility to explain the flare brightness is that the emission is highly anisotropic, as would be expected if the emission region moves relativistically toward us. While only mildly relativistic motion with velocities of  {$\approx$0.5c} are observed inside the nebula \citep{scargle_activity_1969,hester_hubble_2002,melatos_2005}, relativistic motion is expected in the pulsar wind and in the downstream medium behind the wind termination \citep{camus_2009}. Relativistic bulk motion is particularly expected at the ``arch shock'' of the wind termination, which has been proposed as the main site of gamma-ray emission \citep{komissarov_2011}.

The flare emission is expected to result from synchrotron radiation by relativistic electrons and positrons (henceforth referred to together as electrons) \citep{abdo_crab_2011}. A new spectral component emerges in the SED during the flare. The hard photon spectrum ($\gamma \approx 1.3$) of the flaring component implies that most of the electron energy is carried by the highest energy electrons. If the electron \emph{particle} density $n(\epsilon)$ per energy at an energy $\epsilon$ is characterized by a power law $n(\epsilon) \sim \epsilon^{-p}$, the spectral index is $p \le 2 \gamma - 1 \approx 1.6$, in a random magnetic field in which the electrons are isotropically distributed. The energy per logarithmic energy interval, which is proportional to $\epsilon^2 \times n(\epsilon)$, is therefore rising with energy. We note that such a spectrum is also inferred for the radio emitting electrons in the Crab nebula, and more generally in pulsar wind nebulas \citep{gaensel_2006,sironi_2011}. While the radio emission is produced  by a different electron population than the gamma-ray flares, efficient particle acceleration appears to be a common feature in these systems.

It is an interesting question how such a hard electron spectrum is
produced.  Standard diffusive shock acceleration typically results in
spectra with $p \ge 2$ \citep{gallant_1992,kirk_2000}. Even
though it has been shown that harder spectra can be produced in
certain field configurations with low-level turbulence
\citep{kirk_1989,summerlin_2011}, these conditions are not expected at
termination shocks of pulsar winds. Additionally, shock acceleration
appears to be inefficient at highly oblique shocks that are representative
of the pulsar wind termination discontinuity
\citep{ellison_2004,summerlin_2011,sironi_2011b}. One alternative is
that magnetic reconnection in the striped pulsar wind might accelerate
particles \citep{lyubarsky_2003,kirk_2004,yuan_2011,bednarek_2011}.
However, simulations show that reconnection behind the pulsar termination probably does not provide the required electron energies to produce gamma-ray emission \citep{sironi_2011}. Another interesting possibility is that acceleration is occurring directly in the electric field induced by the pulsar, as discussed by \citet{abdo_crab_2011}.

The observation of a peak synchrotron energy of $\approx$380 MeV is among the highest yet seen from astrophysical source today. The observation is surprising as particle acceleration in the presence of synchrotron cooling is expected to limit synchrotron emission to photon energies below $\approx$150 MeV \citep{guilbert_1983,de_jager_gamma-ray_1996,komissarov_2011}. Two solutions to this problem have been proposed recently in this context:
\begin{enumerate}
\item The electric field at the acceleration site is larger in magnitude than the magnetic field. This is generally an unstable state in plasma, as charges will short out the electric field; however, temporarily such a configuration is expected, e.g. in magnetic reconnection events. The gamma-ray emission might occur afterward, when the accelerated electrons enter a region of enhanced magnetic fields \citep{uzdensky_2011,cerutti_2011}. More detailed studies are required to assess whether such a scenario is plausible in the nebula environment and can be sustained for the duration of the flares.
\item The gamma rays are emitted in a region of bulk relativistic motion, and are therefore Doppler boosted toward the observer \citep{komissarov_2011}. For a flow moving directly toward us a Lorentz factor $\gtrsim 2$ is sufficient to accommodate the observed peak energy.
\end{enumerate}

Variations in the Doppler boosting can naturally account for the observed flux variation \citep{lyutikov_2011}.  The observed spectral evolution is compatible with such an interpretation: the energy flux of the emission varies approximately as a power of {$\alpha$ = 3.42 $\pm$ 0.86} with the cutoff energy; a correlation with {$\alpha \approx$ 3} is indeed expected for variations produced by changes in relativistic beaming \citep{lind_1985}. 

The flare brightness, the high frequency of the observed peak of the gamma-ray emission, and the spectral evolution during the flare all suggest the presence of relativistic beaming. We therefore conclude that, independent of the location of the emission region and the physical processes responsible for the flares, the emission region is moving relativistically toward us, and changes in its motion are likely the predominant mechanism responsible for the observed flux variations. Such a kinematic explanation does however not address the issue of how a moving source can be created dynamically and sustained radiatively in the face of strong losses. The Crab Nebula still has much more to teach us.

\acknowledgments

The \textit{\textit{Fermi}} LAT Collaboration acknowledges generous ongoing support
from a number of agencies and institutes that have supported both the
development and the operation of the LAT as well as scientific data analysis.
These include the National Aeronautics and Space Administration and the
Department of Energy in the United States, the Commissariat \`a l'Energie Atomique
and the Centre National de la Recherche Scientifique / Institut National de Physique
Nucl\'eaire et de Physique des Particules in France, the Agenzia Spaziale Italiana
and the Istituto Nazionale di Fisica Nucleare in Italy, the Ministry of Education,
Culture, Sports, Science and Technology (MEXT), High Energy Accelerator Research
Organization (KEK) and Japan Aerospace Exploration Agency (JAXA) in Japan, and
the K.~A.~Wallenberg Foundation, the Swedish Research Council and the
Swedish National Space Board in Sweden. 
Additional support for science analysis during the operations phase is gratefully acknowledged from the Istituto
Nazionale di Astrofisica in Italy and the Centre National
d’Etudes Spatiales in France.
JDS is grateful for funding through the NASA
Applied Information Systems Research program.

\end{document}